\documentclass{aastex63}

\submitjournal{AAS Journals}
\shorttitle{Overmassive brown dwarfs}
\shortauthors{Majidi et al.}
\graphicspath{{./}{figures/}}
\usepackage{amsmath}

\usepackage{hyperref}

\hypersetup{
    colorlinks=true,
    linkcolor=blue,
    filecolor=blue,      
    urlcolor=blue,
    citecolor=blue,
}

\begin{document}

\title{Where to find over-massive brown dwarfs: new benchmark systems for binary evolution}

\correspondingauthor{Dorsa Majidi}
\email{dmajidi@wisc.edu}

\author{Dorsa Majidi}
\affiliation{Department of Astronomy, University of Wisconsin, 475 North Charter Street, Madison, WI 53706 USA}
\affiliation{UCLA Physics \& Astronomy Department, 475 Portola Plaza, Los Angeles, CA, 90095 USA}

\author{John C. Forbes}
\affiliation{Flatiron Institute, 162 5th Avenue, New York, NY, 10010}

\author{Abraham Loeb}
\affiliation{Center for Astrophysics $\vert$ Harvard and Smithsonian, 60 Garden Street, Cambridge, MA, 02138}

\begin{abstract}

Under the right conditions brown dwarfs that gain enough mass late in their lives to cross the hydrogen burning limit will not turn into low-mass stars, but rather remain essentially brown dwarf-like. While these objects, called either beige dwarfs or over-massive brown dwarfs, may exist in principle, it remains unclear exactly how they would form astrophysically. We show that accretion from AGB winds, aided by the wind Roche lobe overflow mechanism, is likely to produce a substantial population of observable overmassive brown dwarfs, though other mechanisms are still plausible. Specifically we predict that sun-like stars born with a massive brown dwarf companion on an orbit with a semi-major axis of order 10 AU will likely produce overmassive brown dwarfs, which may be found today as companions to the donor star's remnant white dwarf. The identification and characterization of such an object would produce unique constraints on binary evolution because there is a solid upper limit on the brown dwarf's initial mass.

\end{abstract}

\section{Introduction} \label{sec:intro}

Brown dwarfs were predicted on theoretical grounds by S. Kumar in the early 1960s \citep{KumarModels, KumarStudyofDegeneracy, KumarHelmholtz, KumarStructureOfStars, KumarBottomOfMainSequence}. It took several decades before searches for brown dwarfs \citep[e.g.][]{1989LathamL} yielded a definitive discovery \citep{Nakajima1995Discovery, 1995Oppenheimer}. In the intervening decades, surveys have identified thousands of brown dwarf candidates, particularly in the infrared and taking advantage of proper motion measurements \citep{1999KirkpatrickDwarfsCooler, 2000LucasAPopulation, 2002GizisPalomar, 2000GizisNewNeighbors, cruz_meeting_2003, 2003Close, faherty_brown_2009, 2011KirkpatrickTheFirstHundred, 2011WestTheSloanDigitalSky}. 

Brown dwarfs have proved enigmatic as the ``missing link'' between planets and low-mass stars,  with some ambiguity remaining about whether brown dwarfs tend to form like scaled-up planets \citep{2009Stamatellos}, or scaled-down stars \citep{2001ReipurthTheFormation, 2002BateTheFormation}. The atmospheres of brown dwarfs have also proven incredibly challenging to model as the result of chemistry and cloud physics \citep{1996Marley, 1997BurrowsNongrayTheory, allard_model_1997, 2011AllardModelAtmospheres, 2012AllardModelsOfVeryLowMass}, which in turn affects their evolution \citep{chabrier_structure_1997, chabrier_theory_2000, 2000ChabrierEvolutionaryModels,2001BurrowsTheTheoryOfBrownDwarfs, baraffe_evolutionary_2003} and hinders the inference of basic properties of field brown dwarfs \citep{filippazzo_fundamental_2015} .

Around the time of their discovery, brown dwarfs and even brown dwarfs exceeding the hydrogen burning limit (HBL) were proposed as constituents of dark matter \citep{salpeter_minimum_1992, hansen_origin_1999, lynden-bell_russell_2001}, but stringent limits on massive compact halo objects from microlensing \citep[e.g.][]{1993AlcockPossibleGravitational, alcock_macho_1997} have made this explanation for dark matter untenable. Nonetheless, the possibility that brown dwarfs more massive than the HBL exist remains intriguing but unrealized thus far observationally. Recently \citep[][hereafter FL19]{forbes_existence_2019} re-discovered this possibility as a consequence of the diagram of equilibrium hydrogen burning as a function of a low-mass star's central entropy and mass (their Figure 1). Brown dwarfs with masses exceeding the HBL\footnote{called variously overmassive brown dwarfs, to emphasize their similarity to brown dwarfs in all aspects but their mass, or beige dwarfs, to emphasize their simultaneous similarity to brown dwarfs and white dwarfs, or pristine white dwarfs, since they may be thought of as hydrogen-rich white dwarfs} need to wait to cool down after their formation, which likely takes several Gyr at least, in order to prevent their cores from igniting when they cross the HBL. In the formulation of FL19, their core entropy needs to be lower than the abscissa of the minimum of the curve in core entropy-mass space that coincides with the HBL.

In this paper, we discuss whether these objects can be produced under specific scenarios, and if so, where observers can find them. We rely heavily on simulations conducted with the Modules for Experiments in Stella Astrophysics (MESA) \citep{2011Paxton,paxton_modules_2013,paxton_modules_2015,paxton_modules_2018,paxton_modules_2019}. In section \ref{sec:First Scenario} we show that the favored scenario from FL19, namely a brown dwarf-brown dwarf binary evolving under gravitational wave radiation to a Roche lobe overflow phase is unlikely to work because of the very small initial semi-major axis required for this scenario. Then in section \ref{sec:creation} we turn to a more-promising scenario, namely wind mass transfer from an evolved donor star. We discuss the challenges and implications of finding over-massive brown dwarfs in section \ref{sec:discussion}, and we conclude in section \ref{sec:conclusion}.

\section{Brown Dwarf Binary} \label{sec:First Scenario}

FL19's preferred scenario for creating over-massive brown dwarfs was a brown dwarf-brown dwarf binary evolving under gravitational wave emission. They argued that over-massive brown dwarfs must be created slowly, i.e. with a very low accretion rate, to avoid igniting the core of the brown dwarf. In addition to being slow, the mass transfer must occur after several Gyr of evolution of the accretor brown dwarf so that the core can become sufficiently degenerate that additional mass would not simply place the star into the ordinary pre-main-sequence region of a diagram in central entropy vs. mass (FL19's Figures 1 and 2). Gravitational wave-driven orbital evolution of a binary brown dwarf system where the accretor had a mass of $\sim 0.07 M_\odot$ and the donor $\lesssim 0.05 M_\odot$ provided a promising avenue for each of these requirements. The mass transfer occurs over several Gyr, and is stable as the result of the mass ratio, $q<0.05/0.07$, being less than the classical critical value of $\sim 5/6
.$ \citep{hjellming_thresholds_1987, soberman_stability_1997}. Additionally, the evolution would not occur immediately, but rather only once the binary had lost enough angular momentum for the donor object to begin overflowing its Roche Lobe. Over several Gyr, this system could transfer $\sim 0.02 M_\odot$ from the donor brown dwarf to the accretor, clearly bringing it from below the HBL to above it.

In order for the Roche Lobe overflow to occur, the binary brown dwarf system needs to end up with a semi-major axis, $a$, comparable to the equivalent Roche lobe radius of the donor object, namely $R_\mathrm{donor} \approx R_\mathrm{L,donor}$, where the Roche lobe radius is given by $R_\mathrm{L,donor}=a G(q)$ with
\begin{equation}
     G(q) = \frac{0.49 q^{2/3}}{0.6q^{2/3} + \ln(1 + q^{1/3})}
\end{equation}
given by the Eggleton formula \citep{eggleton_aproximations_1983}. Here $q = M_\mathrm{donor}/M_\mathrm{accretor}$.

\begin{figure}[ht!]
\plotone{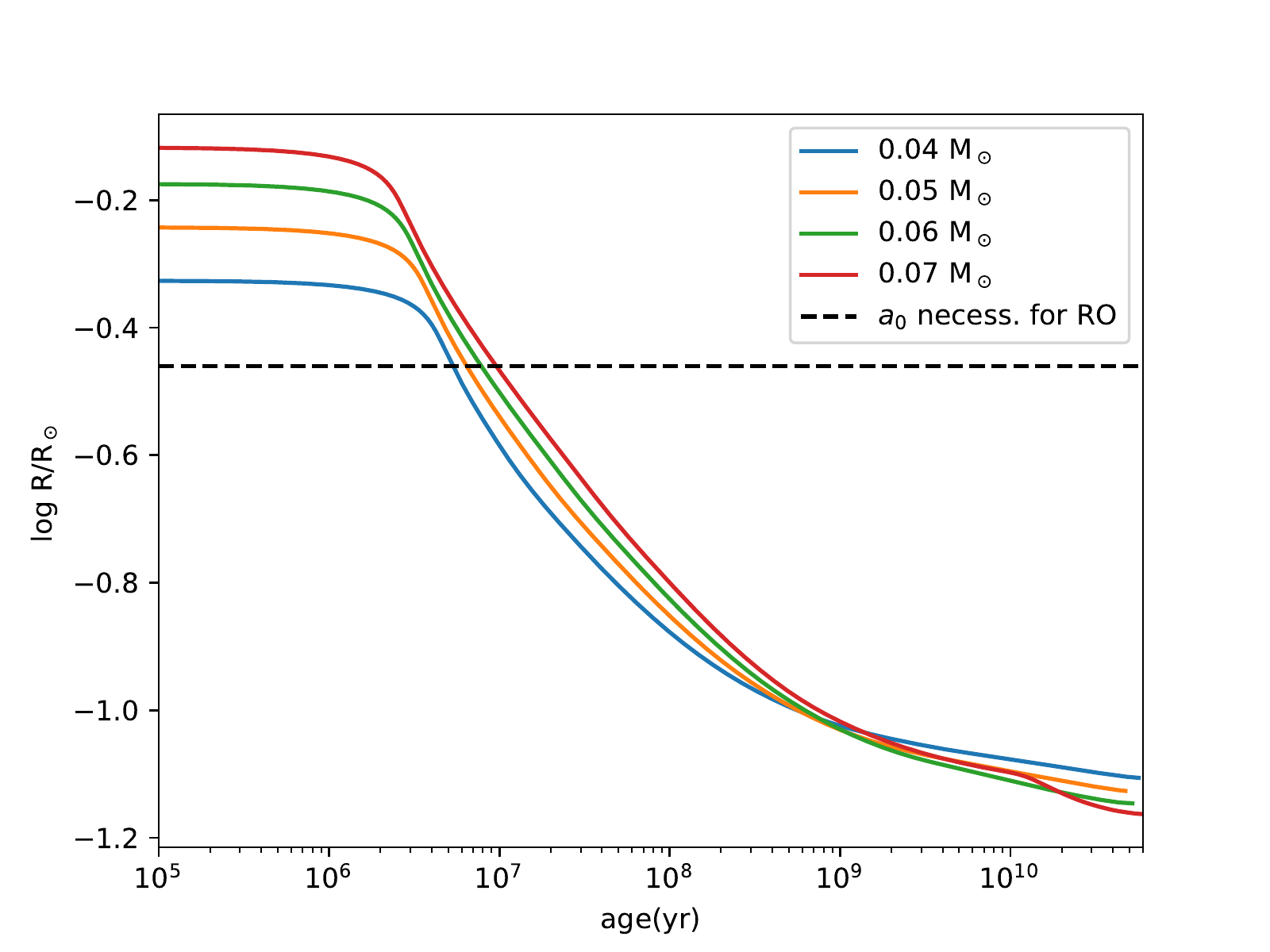}
\caption{This is the plot of the radii as a function of age of a variety of brown dwarfs with different masses. The horizontal dashed line represents the semi-major axis of the brown dwarf-brown dwarf orbit necessary for Roche lobe overflow. \label{fig:general}}
\end{figure}

Under evolution by gravitational wave emission, the initial and final separations, $a_0$ and $a_f$, of a binary are related by
\begin{equation}
\label{eq:a0}
a_0 = \left(a_f^4 + \frac{256 G^3}{5 c^5}M_1 M_2 (M_1 + M_2)t\right)^\frac{1}{4},
\end{equation}
with component masses $M_1$ and $M_2$, and $t$ the amount of time evolved \citep[e.g.][,FL19]{paczynski_gravitational_1967}. For a final orbital separation of $a_f=R_\mathrm{L,donor} \approx 0.19 R_\odot$, assuming $q=5/7$ and $R_\mathrm{donor} = 0.1 R_\odot$, Equation \eqref{eq:a0} implies that the initial separation should be $a_0 \sim 0.35 R_\odot$ for $t=5\ \mathrm{Gyr}$. Comparing this separation to the evolution of a set of brown dwarfs simulated with MESA\footnote{All of the inlists for the MESA runs presented in this work are available at \url{https://github.com/dorsa1376/OMBD}} \citep{2011Paxton} suggests (Figure \ref{fig:general}) that for this scenario to work as imagined in FL19, the orbital evolution dominated by gravitational waves could only begin after the brown dwarfs were $\ga 10^7\ \mathrm{yr}$ old. Any earlier, and the brown dwarfs would collide. Note that the radius evolution depicted here is in good agreement with the recently-discovered and well-characterized transiting brown dwarf TOI-811b \citep{carmichael_toi811b_2021}.

This delay is possible, e.g. if the brown dwarfs are initially separated by a larger distance and lose angular momentum via a different channel, e.g. magnetic braking \citep{rappaport_new_1983}, tidal evolution \citep{hurley_evolution_2002}, Kozai-Lidov oscillations combined with tidal circularization \citep[see][for a review]{naoz_eccentric_2016}, or some other dynamical assembly process. The binary fraction for brown dwarfs is not intrinsically high \citep{fontanive_constraining_2018}, so while the assembly of these systems may be possible, we expect it is not the primary channel by which one might form overmassive brown dwarfs.

\section{AGB binary and creation of an Over-massive Brown Dwarf} \label{sec:creation}

A key motivation for FL19's focus on the binary brown dwarf scenario was the requirement that the accretion proceed slowly. In general, accreting gas will heat the surface of the brown dwarf, which may in turn be transferred to the core. However, it may also be the case that the accretion energy is quickly radiated away \citep{nomoto_rejuvenation_1977, nomoto_accreting_1982, townsley_theoretical_2004}. \citet{paxton_modules_2013} argue that the relevant timescales in a thin layer of mass $\Delta M$ near the surface are the thermal time, $t_\mathrm{th} = C_P T \Delta M / L$, and the accretion timescale $t_\mathrm{acc} = \Delta M / \dot{M}$. To understand whether the accretion luminosity $L_\mathrm{acc} = G M \dot{M}/R$ can be radiated away faster than more mass is added, we need to estimate the dimensionless ratio
\begin{equation}
    \frac{t_\mathrm{th}}{t_\mathrm{acc}} \approx \frac{C_P T R}{G M} \sim 6 \times 10^{-5} \left( \frac{T}{1000 K}\right) \left( \frac{R}{0.1 R_\odot} \right) \left(\frac{0.07 M_\odot}{M}  \right) \ll 1.
\end{equation}
Even if the accretion luminosity can easily be radiated away, additional mass added to the surface of the brown dwarf will require additional pressure support, which can compress and heat the interior. This process is discussed in detail in \citet{townsley_theoretical_2004} in the context of white dwarfs, and implemented numerically in MESA, even in the limit of rapid accretion where $t_\mathrm{th}/t_\mathrm{acc} > 1$.

This means that there may still be a limit to how quickly or how much mass can be added before igniting the brown dwarf. We will return to this point below after seeing the results of a binary MESA run that takes this compressional heating into account. Before that, we need to decide which area of parameter space to focus on. 

The unimportance of the direct accretion luminosity $G M \dot{M}/R$ opens up the possibility of scenarios with much shorter characteristic timescales than the gravitational wave-driven case. These include accretion in AGN disks \citep{cantiello_stellar_2021,jermyn_stellar_2021, dittmann_accretion_2021} and accretion from winds captured gravitationally by a brown dwarf orbiting a stellar companion. Given the uncertainties of stellar evolution in AGN disks and the difficulty of observing brown dwarfs in these distant environments, we focus on the case of wind capture. 

To begin, we consider the case of Bondi-Hoyle accretion \citep{bondi_mechanism_1944, hoyle_effect_1939} from the wind of a star losing mass at a rate $\dot{M}_\mathrm{donor,w}$. The wind has some approximately constant velocity $v_w$, meaning that the density of the wind, $\rho_w$, at any radius away from the star $r$ can be determined by $\dot{M}_\mathrm{donor,w} = 4\pi r^2 \rho_w v_w$. The Bondi-Hoyle formula then implies that the brown dwarf companion accretes at a rate \citep{hurley_evolution_2002}
\begin{equation}
 \epsilon_\mathrm{BH} = \frac{\dot{M}_\mathrm{{BD}}}{-\dot{M}_\mathrm{donor,w}} = \left[\frac{GM_\mathrm{BD}}{v_\mathrm{w}^2}\right]^2 \frac{\alpha_\mathrm{w}}{2a^2} \frac{1}{(1+{v_\mathrm{orb}^2/v_\mathrm{w}^2})^{3/2}} 
 \label{eq:mdotbd}
\end{equation}
where $a$ is the semi-major axis of the orbit, and $v_\mathrm{orb}$ is the orbital velocity $\sqrt{ GM_\mathrm{tot}/a}$ where $M_\mathrm{tot} = M_\mathrm{BD} + M_\mathrm{donor}$. The $\alpha_\mathrm{w}$ parameter is set to $3/2$ following \citet{hurley_evolution_2002} and \citet{boffin_can_1988}. The wind velocity $v_w$ is taken to be proportional to the escape velocity at the surface of the donor star,
\begin{equation}
    v_\mathrm{w}^2 = 2\beta_\mathrm{w} \frac{G M_\mathrm{donor}}{R_\mathrm{donor}}.
\end{equation}
Again following \citet{hurley_evolution_2002} and the wind mass transfer implementation in MESA we set $\beta_\mathrm{w}=1/8$, as appropriate for the cool supergiant winds where most of the mass transfer takes place \citep{kucinskas_circumstellar_1998, lamers_terminal_1995}. Note that the mass rates of change can be negative, i.e. $\dot{M}_\mathrm{donor} \le 0$ as mass is lost from the donor star, but in our case $\dot{M}_\mathrm{{BD}} \ge 0$. For the purposes of the simple calculation in this section, $\dot{M}_\mathrm{donor} = \dot{M}_\mathrm{donor,w}$, i.e. the donor's mass is only changing due to its winds.

As a result of mass transfer between the stars, the angular momentum of the system changes, resulting in the following evolution equation for the semi-major axis,
\begin{equation}
   \frac{\dot{a}}{a} = -\frac{\dot{M}_\mathrm{donor}}{M_\mathrm{tot}} - \left(\frac{2}{M_\mathrm{donor}} + \frac{1}{M_\mathrm{tot}}\right)\dot{M}_\mathrm{BD},
   \label{eq:adot}
\end{equation}
which is adapted from \citet{hurley_evolution_2002} with the simplifying assumption again that $e=0$.

\begin{figure}
    \centering
    \includegraphics{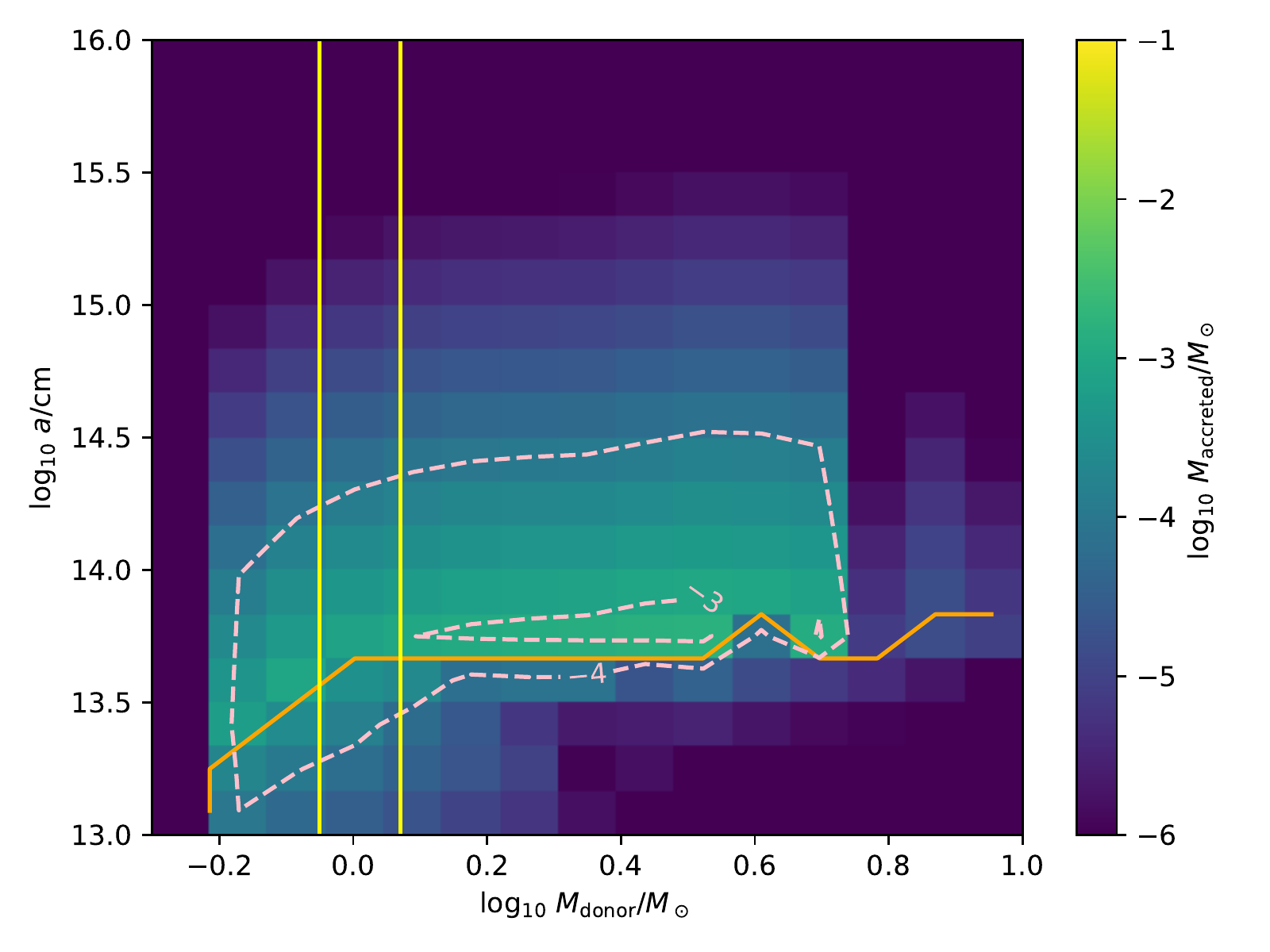}
    \caption{Mass gained by a brown dwarf from wind capture. In this case we assume simple Bondi-Hoyle accretion, and evolve equations \eqref{eq:mdotbd} and \eqref{eq:adot} jointly assuming $R(t)$ and $M(t)$ from the MIST stellar evolution tracks. Systems with separations closer than the orange line undergo a common envelope phase, and systems outside of the yellow lines transfer their mass too early (i.e. the brown dwarf's core is not yet degenerate enough to avoid igniting), or too late (i.e. the donor star's evolution has not had time to occur within the current age of the Universe). The pink contours show values of $M_\mathrm{accreted}$ of $10^{-3} M_\odot$ and $10^{-4} M_\odot$.}
    \label{fig:mgained_mdonor_sep}
\end{figure}

We evolve equations \eqref{eq:mdotbd} and \eqref{eq:adot} jointly, starting from a brown dwarf mass of $0.07 M_\odot$, and a varying initial separation $a(t=0)$. To specify the evolution, we need to know $M_\mathrm{donor}(t)$, and $R_\mathrm{donor}(t)$, which could in principle come from any grid of stellar evolution calculations so long as the brown dwarf companion does not substantially affect the donor star's evolution. Using the MIST tracks \citep{dotter_MESA_2016, choi_mesa_2016}, themselves calculated with a previous version of MESA, we survey the space of initial $M_\mathrm{donor}$ and $a(t=0)$ in Figure \ref{fig:mgained_mdonor_sep}.

In addition to showing $M_\mathrm{accreted}=\int \dot{M}_\mathrm{BD} dt$, i.e. the total mass gained by the brown dwarf over the course of the donor star's evolution, we also mark contours where the initial conditions lead to the brown dwarf being engulfed directly by the donor star at some point during the evolution (light pink lines), or the donor star overflowing its Roche lobe (orange lines), which itself will lead to unstable mass transfer and a common envelope evolution because the mass ratio $q = M_\mathrm{donor}/M_\mathrm{BD} \gg 1$. We also highlight donor masses whose ages correspond to 13.7 Gyr and 3.5 Gyr. Donor stars with longer lifetimes than the former have yet to undergo the red giant and AGB phase where nearly all of the mass transfer takes place, while donor stars with lifetimes shorter than the latter will donate their mass to a brown dwarf that is not yet degenerate enough to avoid ignition of its core\footnote{This boundary is approximate, and has additional dependences on the brown dwarf's initial mass and the quantity of mass actually added (see Figure 2 in FL19), which we will defer to future work.}. This implies that overmassive brown dwarfs formed through this AGB accretion channel live preferentially in a region where the donor star's initial mass is $\sim 1 M_\odot$, and the initial separation of the binary is $\sim 10$ AU. 

To the right of the yellow line, i.e. for donor stars that evolve so quickly that the brown dwarf has not yet cooled, it is likely that objects pushed above the HBL by mass transfer will become low-mass stars. Meanwhile below the orange line, the common envelope phase may itself involve substantial mass transfer or evolution in angular momentum. It is clear that low-mass brown dwarfs can survive a common envelope phase \citep[e.g.][]{maxted_survival_2006, casewell_first_2018}, but it is not clear what would happen to a brown dwarf just below the HBL.

An additional source of uncertainty is the assumption of Bondi-Hoyle accretion. Winds whose velocity is comparable to the escape speed of the system at the Roche lobe set by the binary's potential experience more complex dynamics than is captured by the Bondi problem, as shown in a series of idealized simulations with fixed wind launching velocity and binary configuration \citep{mohamed_wind_2007} in which the accretion rate can exceed the Bondi Hoyle estimate by up to two orders of magnitude. This effect is referred to as the Wind Roche lobe overflow (WRLOF) mode.

These simulation results were fit to a quadratic formula for accretion efficiency in \citet{abate_wind_2013}, namely
\begin{equation}
    \epsilon_\mathrm{WRLOF} = \frac{\dot{M}_\mathrm{accretor}}{-\dot{M}_\mathrm{donor,w}} = \min\left[0.5, \max\left\{\epsilon_\mathrm{BH}, f(q') \left(-0.284 x^2 + 0.918 x -0.234 \right)\right\}\right],
\end{equation}
where $x = R_d / R_\mathrm{L,donor}$, the ratio of the radius where dust forms and hence drives the acceleration of the stellar wind, and the Roche radius of the donor star. The quadratic in $x$ is bound from below by the Bondi-Hoyle efficiency $\epsilon_\mathrm{BH}$ and above by $0.5$, though the exact upper bound is highly uncertain. Following \citet{abate_wind_2013} we adopt a dust radius $R_d$ from \citet{hofner_headwind_2007} of $(1/2) R_\mathrm{donor} (T_\mathrm{eff}/T_\mathrm{cond})^{2.5}$. We adopt a condensation temperature $T_\mathrm{cond}=1000\ \mathrm{K}$ appropriate for oxygen-rich dust with C/O$<1$, and note that now, in addition to $R$ and $\dot{M}_w$, we need to also know the donor star's $T_\mathrm{eff}$ as a function of time. We use $q' = 1/q$ in this formula to account for the opposite mass ratio conventions in Eggleton's and Abate's work. The function $f(q')$ must be 1 for $q'=0.6$, since the \citet{mohamed_wind_2007} simulations are all carried out for $q'=0.6$, but otherwise its functional form must be guessed.

\citet{abate_wind_2013} note that for fixed $M_\mathrm{donor}$, the Bondi Hoyle accretion efficiency $\epsilon_\mathrm{BH} \propto q'^2$ or $q'^2(1+q')^{-3/2}$, depending on the ratio of $v_w$ to $v_\mathrm{orb}$. \citet{abate_wind_2013} argue that $f(q')$ should therefore be approximately bounded by $(25/9)q'^2$ from below and $1$ from above. However, for a brown dwarf orbiting an AGB star, $q' \ll 1$, meaning that in either limit of the Bondi Hoyle accretion efficiency $\epsilon_\mathrm{BH} \propto q'^2$. However, we emphasize that the dynamics of the flow around the brown dwarf could in principle be quite complex, in which case $f$ may not have such a simple dependence on $q'$, and indeed may depend not just on $q'$ but other details of the system. Nonetheless, we adopt $f(q') = (25/9)q'^2$ as a reasonable guess.

\begin{figure}
    \centering
    \includegraphics{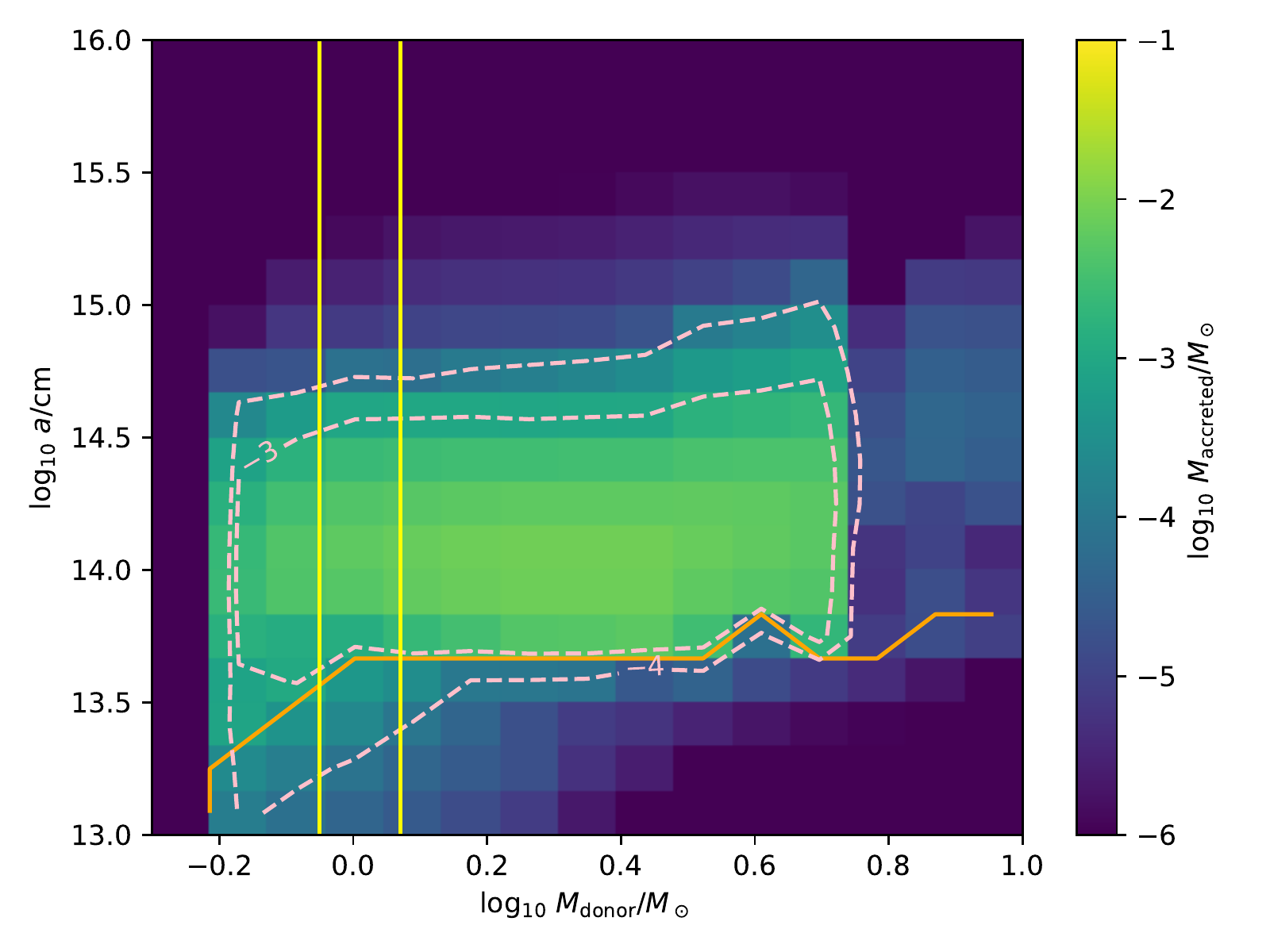}
    \caption{Mass gained by the brown dwarf as a function of the donor star's initial mass and the initial separation of the binary in the WRLOF scenario. Contours and colorbar have the same meaning as in Figure \ref{fig:mgained_mdonor_sep}. In this case, the mass transferred in the AGB phase is substantially larger than in the Bondi Hoyle case, and almost constant at values approaching $10^{-2} M_\odot$, provided that the brown dwarf is initially between about 5 AU and 30 AU and the donor star is initially $\la 6\ M_\odot$. }
    \label{fig:mgained_abate}
\end{figure}

Under these assumptions, we re-run our integrations of $\dot{a}$ and $\dot{M}_\mathrm{BD}$, still using the MIST stellar evolution tracks, resulting in the distribution of $M_\mathrm{accreted}$ shown in Figure \ref{fig:mgained_abate}. As might be expected, substantially more mass is transferred in this WRLOF scenario, with the region expected to be able to produce overmassive brown dwarfs increasing $M_\mathrm{accreted}$ from $\la 10^{-3} M_\odot$ to $\sim 10^{-2} M_\odot$. To understand one of these systems in more depth, we employed MESA's binary evolution capabilities \citep[introduced in][]{paxton_modules_2015} to simultaneously simulate the brown dwarf's and the donor star's evolution, along with the evolution of the binary parameters. This requires enabling the WRLOF accretion efficiency with MESA's run\_binary\_extras feature, and some slight modifications to the default runtime parameters to aid the donor star's passage through the RGB and He flash phases of its evolution. In particular we set the stellar wind coefficients for the \citet{reimers_circumstellar_1975} and \citet{bloecker_stellar_1995} scaling laws to 0.7.  

\begin{figure}[ht!]
\plotone{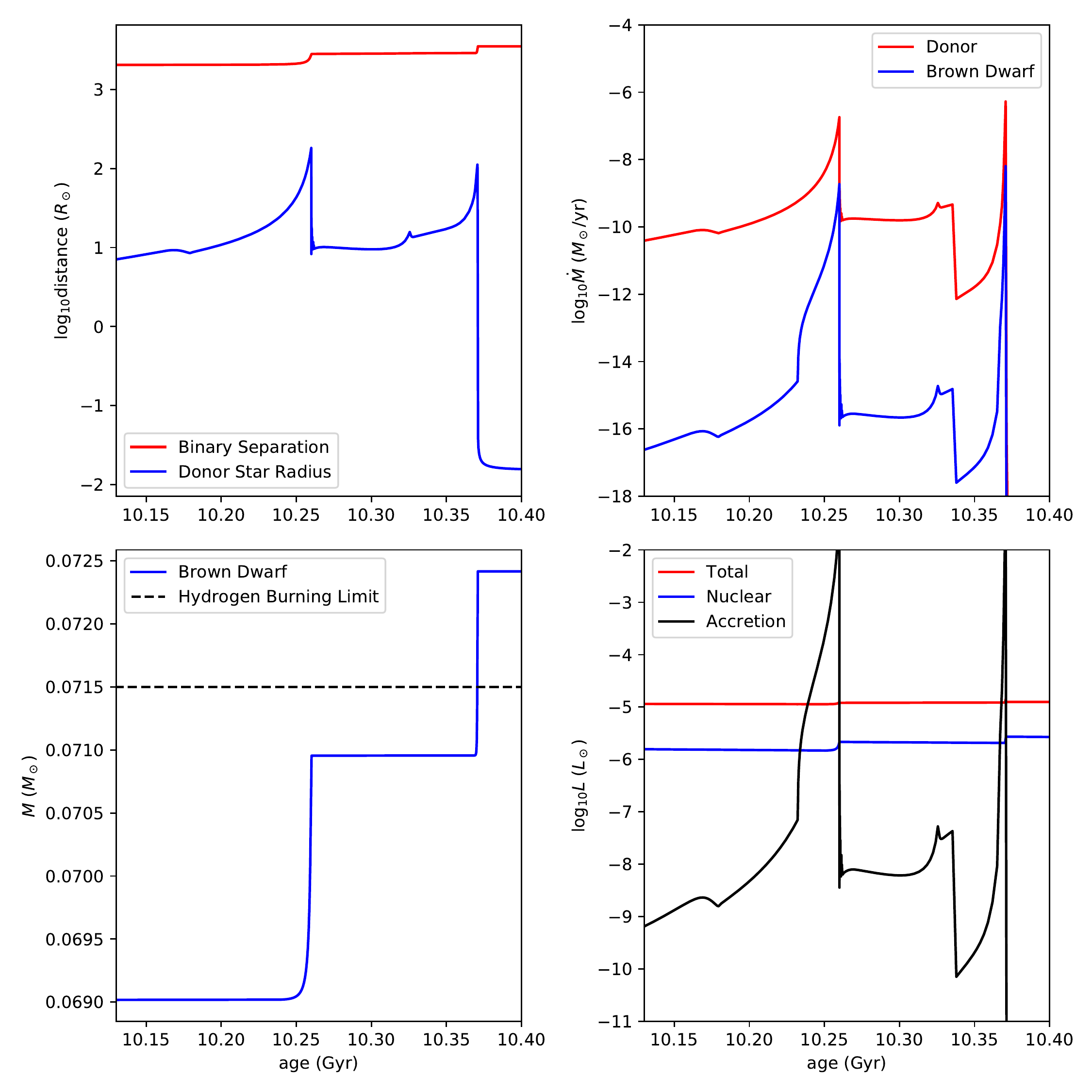}
\caption{AGB binary results. Starting with the upper left plot, the donor star significantly expands its radius at at around 10.25 and 10.37 Gyrs. Yet, the binary separation increases only by a very small distance at the same age. Therefore, we would expect the most mass transfer to happen at those periods. Looking at the lower left panel, we can confirm this idea by the sharp peaks in BD's mass at the same ages of $\sim$10.25 Gyrs and $\sim$10.37 Gyrs. These results are also consistent with the two plots on the right panel. On the upper right plot, the most mass accretion rates are at those ages, and in the lower right plot, the peaks in the accretion, total, and nuclear luminosity are also at the same times. 
\label{fig:AGBresults}}
\end{figure}

\begin{figure}[ht!]
\plotone{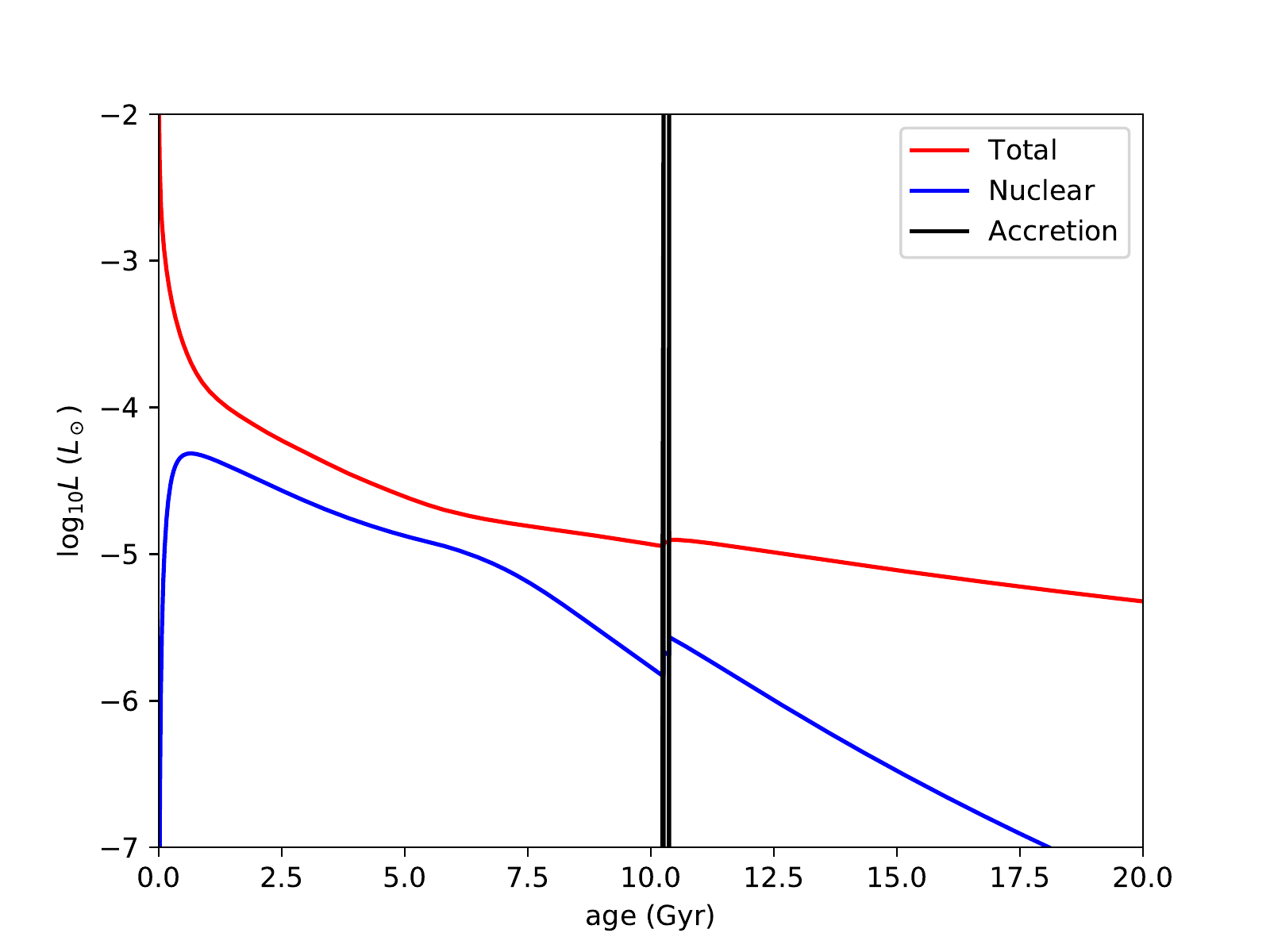}
\caption{BD's luminosity in the long term. We can see that throughout the time span of 20 Gyrs, the BD never ignites to a low-mass star, and the luminosity only decreases after the accretion period at $\sim$10 Gyrs.
\label{fig:20Gyrs}}
\end{figure}

The donor star begins at $1.05 M_\odot$, and the brown dwarf at $0.069 M_\odot$, with an initial period of $10^4$ days, corresponding to an initial semi-major axis of about $9.4$ AU. The evolution of the binary during the vast majority of the mass transfer is shown in Figure \ref{fig:AGBresults}. The mass transfer is dominated by extremely short periods of time where the donor star reaches a radius of around $100 R_\odot$, which correspondingly decreases $v_w$ and dramatically increases the accretion efficiency $\epsilon_\mathrm{WRLOF}$, as evidenced in the decrease in the separation between the wind mass loss rate and the brown dwarf's accretion rate in the upper right panel. The brown dwarf in this scenario clearly crosses the boundary corresponding to the hydrogen burning limit in our version of MESA with a fixed set of atmospheric, opacity, and other physical parameters\footnote{This value of the HBL was determined by running a series of low-mass star/brown dwarf models with the same physics, altering only the initial mass, and zeroing in by-hand with the bisection method by choosing a new mass based on whether a given run lived for $\sim 10$ Gyr before the model cooled (as a brown dwarf) or lived for $\sim 10^{13}$ years before undergoing post main sequence evolution.}. The nuclear luminosity does increase modestly as a result of the new mass, slightly changing the long-run evolution of the brown dwarf's long-run cooling (see Figure \ref{fig:20Gyrs} for the long-run evolution of the brown dwarf's luminosity), but clearly this object remains essentially a brown dwarf in that it never burns hydrogen at a high enough rate to supply its surface luminosity.

The MESA run's sudden increase in nuclear luminosity as a result of the mass transfer from the donor star's wind strongly suggests that there are limits to how much mass can be transferred before the brown dwarf, now above the hydrogen burning limit, has its core ``jump-started'' to produce enough nuclear luminosity to supply the surface luminosity, thereby becoming an ``ordinary'' low-mass star, albeit with an unusual history. We find that within a particular short period of accretion, i.e. the AGB or RGB phase of the donor star, the brown dwarf's central density and temperature both increase. In particular, $d ln T_c/dt \approx d \ln M/dt$ and $d ln \rho_c/dt \approx 2 d \ln M/dt$. This is not dissimilar to the case of compressional heating in a white dwarf, for which \citet{townsley_theoretical_2004} showed that in the isothermal core, compressional heating follows $d \ln T_c/dt \approx (6.4/a) d \ln M/dt$ where $a$ is a constant of order unity related to the structure of the white dwarf's core. We defer an exact accounting of the maximum mass obtainable by an overmassive brown dwarf before it becomes a star to future work.

While the binary MESA run is promising, in that it shows how the entire binary system evolves together to allow the brown dwarf to exceed the HBL without igniting as a low-mass star, including the effects of compressional heating, we note that there is a discrepancy between this particular MESA run and the corresponding integration we carried out using the MIST stellar tracks. Rather than gaining a mass of $0.0034 M_\odot$, the integration predicts about twice the mass gain, $0.0066 M_\odot$. This difference appears to arise from a difference in the evolutionary path of the $1.1 M_\odot$ star according to MIST and the path taken by the present MESA simulation. In the MIST track, the star loses a much higher proportion of its mass in an AGB phase where it reaches a much larger radius. This is likely the result of the higher wind scaling factors we employed in the present run to allow it to ``survive'' the helium flash without crashing. We expect that, especially for properties like $T_\mathrm{eff}(t)$, $R(t)$, and $\dot{M}(t)$ that are observable with only limited model-dependence, the MIST track is more realistic, since the parameters were tuned to match a variety of observations.

\begin{figure}
    \centering
    \includegraphics{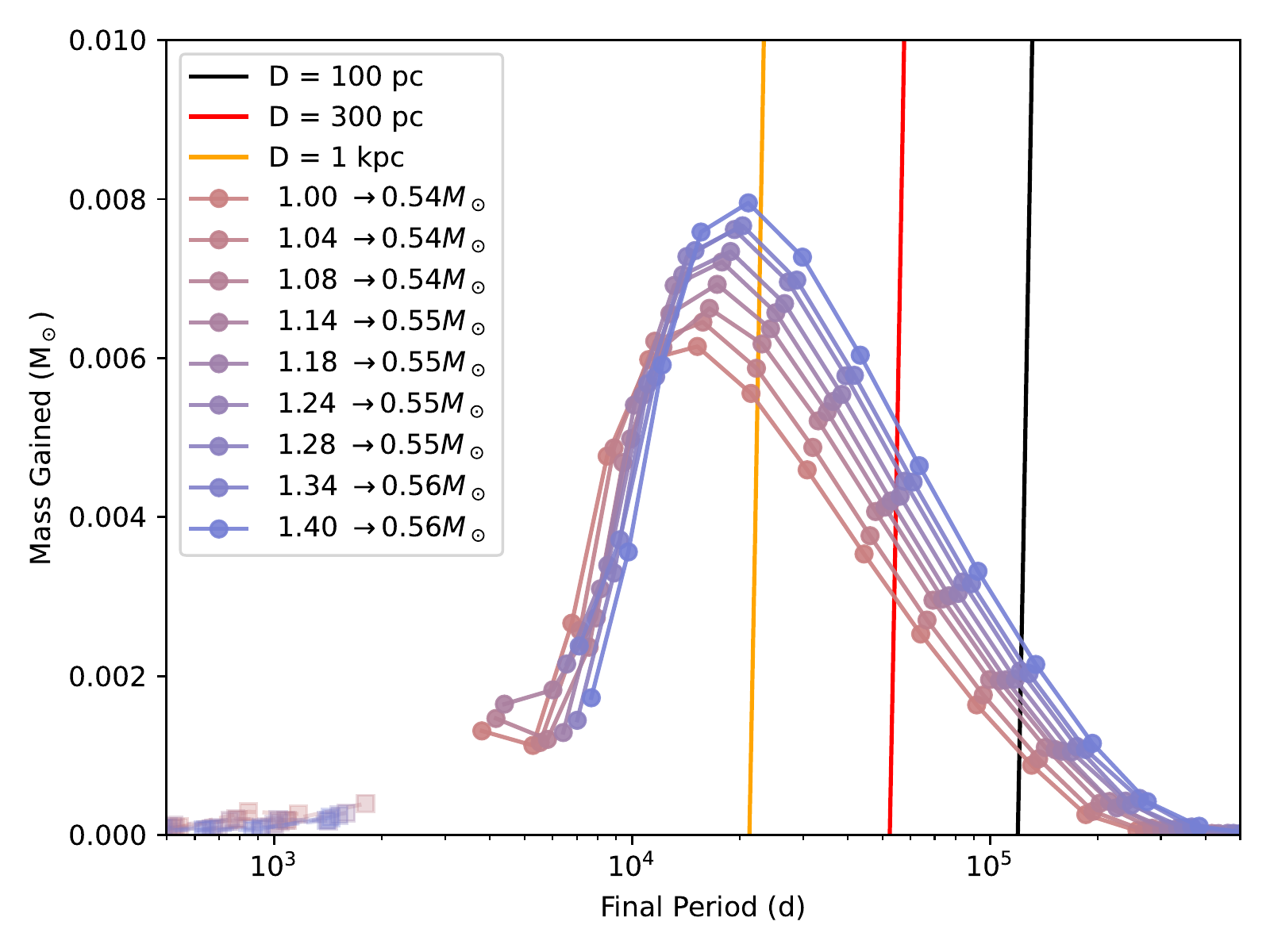}
    \caption{Mass gained by a brown dwarf as a function of the period of the resulting white dwarf - brown dwarf binary. Each red-blue curve corresponds to a different main sequence mass of the donor star, with the legend showing these starting masses and the final mass of the white dwarf at the end of the MIST track. The different points correspond to different initial separations between the donor star and the brown dwarf, i.e. the same set of integrations shown in Figure \ref{fig:mgained_abate}. Square points on the left side of the plot arise from integrations where the brown dwarf is at one point enveloped by the donor star. The nearly-vertical lines show the locus of final periods and masses gained where Gaia is expected to be able to measure the mass of the brown dwarf to $10\%$ given only its astrometric observations of the white dwarf. The most promising systems for solidly discovering overmassive brown dwarfs (with Gaia) are therefore white dwarf-brown dwarf binaries, within 300 pc -1 kpc distance from Earth, whose orbital period today is about 30-90 years.}
    \label{fig:mgained_final}
\end{figure}

\section{Discussion} \label{sec:discussion}

Having found the conditions under which overmassive brown dwarfs may form, we now turn to the questions of how common they may be, how they may be detected, and why they may be of interest beyond their novelty.

\subsection{Occurrence Rate}

Brown dwarf companions of sun-like stars, i.e. the putative progenitors of overmassive brown dwarfs, are known to be rare, an observation referred to as the brown dwarf desert \citep{marcy_planets_2000, 
grether_how_2006, raghavan_survey_2010}. While there is a particular lack of brown dwarfs on short orbits, as demonstrated in radial velocity searches, direct imaging and interferometric searches \citep[e.g.][]{gizis_substellar_2001, metchev_palomar_2009,  tanner_highcontrast_2010, kraus_mapping_2011, evans_mapping_2012}, and much longer-baseline radial velocity searches \citep[e.g.][]{kiefer_detection_2019} find a somewhat higher fraction of stars to have brown dwarf companions. Taken together, these results suggest that of order 1\% of sun-like stars appear to have a brown dwarf companion between $\sim 5$ and $\sim 30$ AU. Gaia is expected to dramatically reduce the uncertainty of this fraction by greatly expanding the sample of brown dwarfs around sun-like stars \citep{holl_gaia_2021}. The mass function of companions to sun-like stars is fairly steep \citep[e.g.][]{grether_how_2006}, implying that the brown dwarf companions in question will be preferentially close to the hydrogen burning limit (as opposed to the middle of the brown dwarf mass range), meaning that comparatively little mass transfer may be necessary to bring the brown dwarfs over the HBL into the overmassive brown dwarf regime. Despite the brown dwarf desert, therefore, we expect that only somewhat less than $\sim 1\%$ of white dwarfs whose progenitors are sun-like stars to potentially have an overmassive brown dwarf companion.

The key uncertainty, moreso than the statistics of the population of the progenitor systems, is the physics of the WRLOF mechanism. Setting aside uncertainties in the original \citet{mohamed_wind_2007} simulations themselves, our results are extremely sensitive to guesses made by \citet{abate_wind_2013} for how this mechanism should scale to mass ratios far from the original $q'=0.6$. This provides an interesting opportunity for hydrodynamic simulations of binary evolution \citep[e.g.][]{macleod_bound_2018} to make predictions in the extreme mass ratio regime of the WRLOF mechanism which may be confirmed or falsified with the future detection of overmassive brown dwarfs.

\subsection{Search Methods}

While the statistics of sun-like stars on the main sequence and their companions can tell us a great deal about the frequency with which we may hope to find overmassive brown dwarfs, actually finding them requires seeking out the companions of the white dwarfs left behind after the evolution of a sun-like star through the AGB phase. The long period of the orbits, faint intrinsic brightness of the primaries, and small radii of the primaries make detection via transits and radial velocity methods difficult. To date, many low-mass companions to white dwarfs have been detected via multi-band photometric searches \citep{steele_white_2011, girven_white_2011, rebassa-mansergas_infraredexcess_2019} that take advantage of the white dwarf's and brown dwarf's dramatically different spectra, peaking in the UV/blue and IR respectively. These candidate systems may then be followed up with direct imaging, which benefits from the reduced contrast between the two objects relative to the main sequence star - brown dwarf case, and spectroscopy \citep[e.g.][]{hogg_confirming_2020} to characterize at least the nature of the companion object, though perhaps not the orbit. Timing variability in pulsating white dwarfs may have also been used to find candidate companions \citep{mullally_substellar_2007,winget_limits_2015}. Intriguingly of the known white dwarf binary systems, few have been reported to have brown dwarf companions, but there are some objects with masses just above the HBL \citep{farihi_lowluminosity_2005} (though we caution that the steep mass function of companions in this regime and the spectral-type classification used to estimate these masses makes it a priori unlikely that these objects are overmassive brown dwarfs).

Astrometry is also an extremely powerful tool for identifying and characterizing systems in this regime \citep{sahlmann_astrometric_2013, bowler_orbit_2018, dieterich_dynamical_2018}, with companion masses expected to be reasonably accurately measured in the brown dwarf regime out to hundreds of pc \citep{andrews_weighing_2019} with Gaia. Gaia is expected to dramatically expand the number of known white dwarfs and white dwarf binaries \citep{toonen_binarity_2017}, particularly in future data releases that include or allow explicit binary modeling. 

\subsection{Identification of Overmassive Brown Dwarfs}

A key obstacle to the discovery of an overmassive brown dwarf is positively distinguishing it from other similar objects, in particular low-mass stars and ``ordinary'' brown dwarfs. Roughly speaking, an overmassive brown dwarf at the same mass as a low-mass star will be far more compact, cooler, and dimmer (see figures 3 and 4 of FL19). Meanwhile the differences between an overmassive brown dwarf and an ordinary brown dwarf are quite modest. This emphasizes the importance of a precise mass measurement through either astrometry or radial velocity measurements \citep[or their combination, e.g.][]{brandt_precise_2019}, both to ensure that the object is above the HBL and by how much. Figure \ref{fig:mgained_final} shows the result of our orbital integrations with the MIST stellar tracks for the final period of the overmassive brown dwarf - white dwarf binary that is the end state of these systems. Their $\gtrsim 30$-year periods mean that precisely characterizing these systems by radial velocities will be difficult in general, especially given that white dwarfs are not commonly included in radial velocity campaigns owing to their broad lines and intrinsic faintness.

Gaia astrometry seems to be a more promising approach: \citet{andrews_weighing_2021} determine a simulation-based empirical estimate for the precision in mass that Gaia may achieve over the course of its lifetime, even for orbital periods far exceeding the time baseline of 5-10 years expected for the observations. Essentially the deviation, induced by an unseen companion, on a given object from the helical pattern it traces in the plane of the sky may be measured even if only a small fraction of a single arc of the helix is traced in Gaia's lifetime. Their result for the fractional precision $\xi$ that may be obtained is
\begin{equation}
\xi \approx 35 \frac{D R}{V^2}\frac{\sigma_G}{\tau_G^2}\sqrt\frac1N,    
\end{equation}
where the distance to the system from Earth is $D$, $R$ is the radius of the orbit of the observed object, i.e. the white dwarf in this case, and similarly $V$ is the orbital velocity of the white dwarf. The lifetime of Gaia is $\tau_G$, its precision is $\sigma_G$, and $N$ is the number of observations of the system. Note that $\xi$, $\sigma_G$ when expressed in radians, and of course N, are dimensionless. Following \citet{andrews_weighing_2021}, we adopt $N=70$, $\tau_G=5$ years, and $\sigma_G = 10 \mu as$. For the orbital velocity and radius, we use standard solutions for the two-body problem for circular orbits to find the white dwarf's orbital velocity and radius - in so doing we fix the mass of the white dwarf to $0.55 M_\odot$, which is typical of the final masses of the white dwarfs for this range of donor masses according to the MIST tracks. We also retain the dependence on the mass gained by the brown dwarf. We then fix $D$ to three different values, and plot (Figure \ref{fig:mgained_final}) the resulting curves, which on this scale are nearly straight vertical lines, where $\xi = 0.1$. This implies that Gaia should be able to determine the mass of these brown dwarf companions with sufficient precision, out to hundreds of parsecs, to assess whether they exceed the HBL or not. While this is a remarkable and encouraging testament to the power of Gaia, there are several additional caveats that bear consideration.

The HBL itself is uncertain \citep[for a useful compilation see e.g. Table 5 of][]{dieterich_dynamical_2018}. While we can determine what the HBL is for a given set of physics in a particular version of MESA by trial and error, the complex chemistry and cloud physics in the atmospheres of brown dwarfs renders these determinations unreliable. Empirical determinations of the HBL by characterizing transiting systems with a component near the HBL \citep[e.g.][]{grieves_populating_2021} is a promising avenue forward. An additional complication is that the objects in question may have non-solar metallicities especially given the old age required to produce overmassive brown dwarfs. To a first approximation, lower metallicities push the HBL to higher masses, meaning that even if an overmassive brown dwarf candidate has a precisely-measured mass securely above the solar-metallicity HBL, abundance measurements will need to be made to determine whether the object is a relatively high-metallicity overmassive brown dwarf, or a low-metallicity ordinary brown dwarf.

\subsection{What could we learn from a discovery?}

Discovering a bona fide overmassive brown dwarf would be interesting in and of itself, vindicating the fundamental physics of hydrogen burning in degenerate matter and the evolution of stellar structure under accretion. It would also directly verify the WRLOF mechanism without which brown dwarfs would have a difficult time accreting enough mass from their AGB companions to exceed the HBL. More generally, an overmassive brown dwarf would provide a completely unique constraint on the process of binary mass transfer because an overmassive brown dwarf {\it must} start with a mass below the HBL, meaning that 
\begin{equation}
M_\mathrm{accreted} > M_\mathrm{beige} - M_\mathrm{HBL},
\end{equation}
where $M_\mathrm{beige}$ is the mass of the overmassive brown dwarf at the time of observation. As we have seen in our joint orbit and mass transfer integrations $M_\mathrm{accreted}$ is sensitive to the wind prescription, which not only sets the instantaneous mass loss rate, but also the subsequent evolution of the donor star, determining how much mass it loses in different phases of its evolution, and how much it inflates during the AGB phase (see \cite{willson_mass_2000}, and \cite{smith_mass_2014} for a review of uncertainties in the mass loss phase of AGB stars. The evolution of these extremely bright stars is critical in interpreting the SEDs of distant galaxies (see e.g. \cite{maraston_evolutionary_2005} and \cite{choi_mesa_2016}), and may affect the location and size of the putative black hole mass gap \citep{farmer_mind_2019}, the separation in mass scales between low-mass black holes observed electromagnetically and more massive binary mergers seen in gravitational waves.

In addition to these constraints, an overmassive brown dwarf would provide a new approach to constraining the initial mass-final mass relation, i.e. the mapping between a star's zero age main sequence mass and its remnant mass at the end of stellar evolution \citep{kalirai_initialfinal_2008, catalan_initialfinal_2008, andrews_constraints_2015}. Typically this is done by estimating a white dwarf's mass and the time it has spent cooling, and comparing to an independent estimate for the white dwarf's age. The discovery of an overmassive brown dwarf companion to a white dwarf would independently restrict the white dwarf's progenitor to a relatively narrow range of initial masses.

\section{Conclusion}
\label{sec:conclusion}

Overmassive brown dwarfs, or beige dwarfs are a proposed but not-yet-discovered type of stellar object that may form when an old brown dwarf gains enough mass to exceed the HBL without igniting its core. This requires that the core be sufficiently cold and degenerate to sustain some compression and heating by the addition of new mass. We have reviewed several astrophysical mechanisms by which overmassive brown dwarfs may be formed, with the key consideration being the long timescale required to allow the brown dwarf to cool and become degenerate. 

A brown dwarf-brown dwarf binary evolving by gravitational wave emission as proposed by FL19 remains an intriguing possibility for producing overmassive brown dwarfs, but would require some fine-tuning in that the pair of brown dwarfs would need to begin the gravitational wave-mediated inspiral at least about 100 Myr after their formation to avoid an outright collision. While this may be arranged by having the brown dwarfs form at a wider separation and having some other mechanism act to bring them close later in their evolution, this should be a rare occurrence.

We also considered the case of mass transfer from the slow winds of an AGB star being swept up by a brown dwarf companion. Since nearly all of the mass transfer occurs at the end of the donor star's life, there is sufficient time for the brown dwarf to cool before the mass transfer takes place. The requirement that the brown dwarf is allowed to cool and that the donor star must evolve within the age of the Universe specifies a narrow range in donor star masses around $1 M_\odot$ where this mechanism can form overmassive brown dwarfs.

We find that mass is transferred to the brown dwarf most efficiently when its initial semi-major axis around the donor star is around 10 AU. For initial separations within a few AU, the brown dwarf will likely enter a common envelope phase, the result of which we do not attempt to model. For separations beyond about 30 AU, the wind becomes low enough density that relatively little mass is transferred.

Taken together, these restrictions single out a very particular type of progenitor system, i.e. a $1 M_\odot$ star with a brown dwarf companion at $\sim 10$ AU, which may result in an overmassive brown dwarf. This overmassive brown dwarf would orbit the white dwarf remnant of the donor star. Candidates for these systems can be identified photometrically taking advantage of the vastly different SEDs of the white dwarf and its potentially overmassive brown dwarf companion. Characterizing the system well enough to be certain that the brown dwarf is overmassive would require followup observations, the most promising of which is astrometric. Future data releases from Gaia may therefore allow overmassive brown dwarfs to be positively identified and characterized.

Such a discovery would provide a unique constraint on stellar winds and mass transfer in binaries. Typical constraints on binary evolution require population synthesis approaches \citep[e.g.][]{moe_mind_2017,mandel_extracting_2019,broekgaarden_stroopwafel_2019,breivik_COSMIC_2020} to construct statistical properties of a synthetic sample of binary progenitor and remnant systems, to be compared with observed populations. In contrast, the discovery of overmassive brown dwarfs could provide meaningful constraints from a single system, because, not only would the current state of the system be characterized, but its state prior to the mass transfer is also constrained: the now-overmassive brown dwarf has to have been below the HBL when the system formed, and the donor star, now a white dwarf, had to have been around $1\ M_\odot$ on the main sequence.

\acknowledgements
This work began as a project at the Banneker Institute at the Center for Astrophysics $\mid$ Harvard \& Smithsonian. We thank the students, mentors, and staff of Banneker, the Institute for Theory and Computation, and the Flatiron Institute for their support. We would also like to thank Adam Jermyn, Katie Breivik, Morgan MacLeod, Smadar Naoz, Brad Hansen, Rosanne Di Stefano, Kevin Schlaufman, Meng Sun, and Selma de Mink for helpful conversations, and the anonymous referee for their constructive report.

\bibliography{OMBDExports.bib,ombd2.bib}

\end{document}